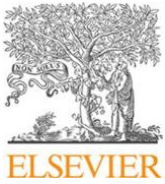

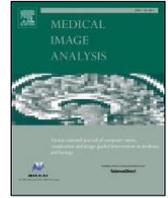

# Self-transfer learning via patches: A prostate cancer triage approach based on bi-parametric MRI

Alvaro Fernandez-Quilez[a,b,f,*], Trygve Eftestøl[c], Morten Goodwin[d], Svein Kjosavik[e], Thor Ole Gulsrud[a], Ketil Oppedal[b,f]

[a]Department of Quality and Health Technology, University of Stavanger, Norway.
[b]Stavanger Medical Imaging Laboratory (SMIL), Department of Radiology, Stavanger University Hospital, Norway.
[c]Department of Electrical Engineering and Computer Science, University of Stavanger, Norway.
[d]Centre for Artificial Intelligence Research (CAIR), Department of ICT, University of Agder, Grimstad, Norway.
[e]General Practice and Care Coordination Research Group, Stavanger University Hospital, Norway.
[f]Centre for Age-Related Medicine, Stavanger University Hospital, Norway.

## ARTICLE INFO

## ABSTRACT

Prostate cancer (PCa) is the second most common cancer diagnosed among men worldwide. Current PCa diagnostic pathway comes at the cost of substantial overdiagnosis, leading to unnecessary treatment and further testing. Bi-parametric magnetic resonance imaging (bp-MRI) based on apparent diffusion coefficient maps (ADC) and T2-weighted (T2w) sequences has been proposed as a triage test to differentiate between clinically significant (cS) and non-clinically significant (ncS) prostate lesions. However, analysis of the sequences relies on expertise, requires specialized training and suffers from inter-observer variability. Deep learning (DL) techniques hold promise in tasks such as classification and detection. Nevertheless, they rely on large amounts of annotated data which is not common in the medical field. In order to palliate such issues, existing works rely on transfer learning (TL) and ImageNet pre-training, which has been proven to be sub-optimal for the medical imaging domain. In this paper, we present a patch-based pre-training strategy to distinguish between cS and ncS lesions which exploits the region of interest (ROI) of the patched source domain to efficiently train a classifier in the full-slice target domain which does not require annotations by making use of transfer learning (TL). We provide a comprehensive comparison between several CNN's architectures and different settings which are presented as a baseline. Moreover, we explore cross-domain TL which exploits both MRI modalities and improves single modality results. Finally, we show how our approaches outperform the standard approaches by a considerable margin.

## 1. Introduction

Prostate cancer (PCa) is the second most commonly diagnosed cancer (Siegel et al., 2020), with an estimated incidence of 1.3 million new cases among men worldwide in 2018 (Bray et al., 2018; Fernandez-Quilez et al., 2020).

Traditionally, diagnosis of PCa has been based on digital rectum examination (DRA). However, ever since measurement of prostate-specific antigen (PSA) levels in serum (PSA testing) was approved as a screening test in the early 1990s, it became the main tool for PCa diagnosis and management (Catalona et al., 1994). Standard diagnostic pathway in men with abnormal PSA levels consisted of transrectal ultrasound (TRUS)-guided biopsies (Hodge et al., 1989). Nevertheless, PSA test-ing comes at the cost of substantial overdiagnosis, which leads to unnecessary biopsies and over-treatment of indolent or low-malignant potential tumors (ncS) whilst underestimating potentially lethal tumors (cS) (Barry, 2009) making stratification of lesions a critical part of the PCa diagnostic pathway (Johnson et al., 2014).

*Corresponding author: email.: alvaro.f.quilez@uis.no;

Thanks to recent advances in image acquisition and interpretation, magnetic resonance imaging (MRI) has proven to be a valuable tool for PCa detection, staging, treatment planning and intervention (Shukla-Dave and Hricak, 2014). In particular, multi-parametric MRI (mp-MRI) has been proposed as a triage test before prostate biopsy due to its detection and diagnostic potential(Dirix et al., 2019; Lomas and Ahmed, 2020).

Although many centers follow a similar mp-MRI acquisition protocol, there is no standardized procedure yet (Yoo et al., 2015). In spite of it, mp-MRI can be defined as the combination of several of the following MRI modalities: T2-weighted (T2w), diffusion weighted (DW), dynamic contrast enhanced (DCE) and MR spectroscopy (MRS) images (Demirel and Davis, 2018).

Prostate Imaging-Reporting and Data System (PI-RADS) was developed in 2013 in an effort to standardize mp-MRI usage in PCa, and updated in 2015 (PI-RADS v2) and 2019 (PI-RADS v2.1) (Weinreb et al., 2016; Turkbey et al., 2019). The latest update relegated DCE to a minor clarification role secondary to T2w and DW, offering the possibility of limiting the triage to non-contrast exams and a bi-parametric setting (bp-MRI), consisting of T2w and DW (and derived sequences from it, Apparent Diffusion Coefficient maps - ADC) sequences (De Visschere et al., 2017). Nevertheless, bp-MRI analysis relies on radiologists' experience and requires specialized training (Gaziev et al., 2016). Moreover, bp-MRI analysis can be a time-intensive task and they are not exempt of inter-observer variability issues, which have been found in the PI-RADS v2.1 protocol (Rosenkrantz et al., 2016).

Deep learning (DL) techniques have led to important breakthroughs in the computer vision area. In particular, convolutional neural networks (CNNs) have been successful using natural images in tasks such as classification (Krizhevsky et al., 2017; Simonyan and Zisserman, 2014). Promise has also been shown in different medical imaging tasks such as classification and detection of tumors in MRI (Cao et al., 2019). Nevertheless, CNNs success in medical imaging is hampered by small amounts of labeled data, imbalance of the data and the highly heterogeneous nature of it (Shin et al., 2016; Esteva et al., 2017).

Several approaches have been proposed to deal with the shortage of data in the medical field. Augmentation techniques (e.g. rotate the image), generation of new images based on generative adversarial networks (GAN) to enlarge the dataset (Fernandez-Quilez et al., 2021), patch-based strategies or transfer learning (TL), being the last one almost an integral part of medical imaging applications. The present standard for TL is to take an existing architecture trained on a large corpus of natural images such as ImageNet (Deng et al., 2009) (source domain) and then fine-tune the model on the limited medical imaging data (target domain) (Rajpurkar et al., 2017). However, such approach has been proven to be sub-optimal for the medical imaging domain, due to the existing gap between ImageNet domain and medical imaging domain. (Raghu et al., 2019).

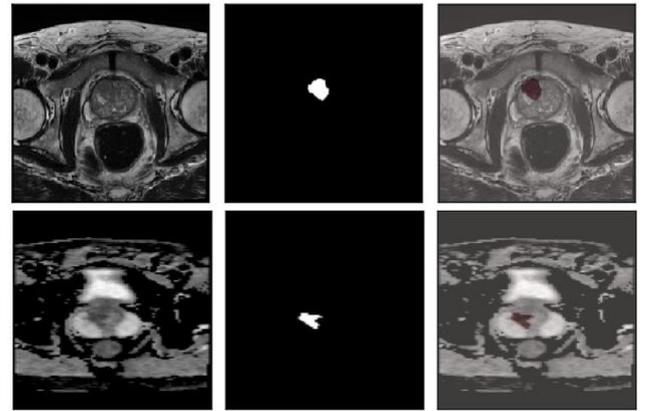

Fig. 1: Examples of T2w and ADC slices and their respective masks (first and second row, respectively) and the result of superimposing the original slice on the mask (final column).

Patch-based strategies refer to the process of dividing an original image into small blocks (patches) in a sliding window fashion. The extracted patches can be extracted in an overlapping or non-overlapping way. Resorting to such strategy allows to increase the amount of data whilst enabling the network to focus on smaller regions of the input, which has been argued to be useful for tumor detection and classification tasks (Shen et al., 2019; Mehrtash et al., 2017). Nevertheless, most studies rely in the ability to successfully extract patches based on the region of interest (ROI) to train an end-to-end classification system, which requires a fully annotated data set or an additional step to detect such ROI, adding extra complexity to the problem at hand and making the usage of the system more difficult due to the labeled data requirements.

### 1.1. Related work

Work found in the literature aiming to stratify PCa lesions can be classified depending on the type of input. As such, some of the work attempt to achieve the classification in cS and ncS categories by making use of MRI slices (2D) while others try to achieve it at the patient level (3D). In both cases, the problem can be approached from a whole image (sequence) point of view or from a 2D or 3D patch-based one. In all the cases works evaluate the quality of the results based on area under the curve (AUC), a standard classification metric. A summary of the related work and results can be found in Table 1.

In both Wang et al. (2017) and Ishioka et al. (2018) they present a work at the slice-level (2D) which makes use of the full slice. They use a single modality (T2w) to achieve an AUC of 0.79 on the test set whilst in the second one they obtain an area under the curve (AUC) of 0.84. In both cases, they make use of a VGG-inspired architecture (Simonyan and Zisserman, 2014). Nevertheless, the work presented a limited evaluation protocol, including only 17 slices to test the algorithm in the first case.

An approach exploiting bp-MRI 2D patches is presented in Le et al. (2017). In particular, they make use of ADC and T2w slices and extract patches based on the ROI. They design a new similarity loss function to fuse both modalities and endorse consistent feature extraction. They obtain an AUC of 0.91 with the best setting and after fusing the information. In spite of the good results based on AUC, the work depends on hand-crafted features, making the replication and generalization of the results hard to achieve (Traverso et al., 2018). In addition, those results might not be representative enough of the reality since the different sets for testing were heavily augmented with the aim of balancing the classes. Following the same type of approach, Liu et al. (2017) presents a mp-MRI approach exploiting DW, T2w, ADC maps and pharmacokinetic parameter ($K^{trans}$). They extract 32x32 2D patches around the ROI and train a VGG-inspired model. They obtain an AUC of 0.84. In the work of Yuan et al. (2019) they present an approach based on mp-MRI (multi-view T2w and ADC). Patches of different sizes are extracted from the different sequences and used as an input for a pre-trained model. Following, a new similarity constraint is imposed to obtain features with intra-class compactness and inter-class separability. Final AUC on the test set is of 0.89. In all the cases, the works heavily rely on ROI annotations at test time, which are rarely available and makes it hard to make use of other institutions data sets without annotations to validate the results.

Table 1: Summary of related work.

| Modality | Authors | Approach | ROI | AUC[t] |
|---|---|---|---|---|
| T2w | (Ishioka et al., 2018) | 2D (slice) | C | 0.84 |
| T2w | (Wang et al., 2017) | 2D (slice) | C | 0.79 |
| T2w and ADC | (Le et al., 2017) | 2D (patch) | √ | 0.91 |
| T2w, DW, ADC and $K^{trans}$ | (Liu et al., 2017) | 2D (patch) | √ | 0.84 |
| multi-view T2w and ADC | (Yuan et al., 2019) | 2D (patch) | √ | 0.89 |
| T2w, DW and DCE | (Mehrtash et al., 2017) | 3D (patch) | √ | 0.80 |
| DW | (Yoo et al., 2019) | 2D and 3D (patch) | √ | 0.87 |
| T2w, ADC and DCE (PI-RADS v2.0) | Radiologist[*] | Manual (Visual) | C | 0.83 |

[*] Radiologist-level AUC (human performance) based on (Mehrtash et al., 2017).
[t] Test AUC.

In the work of Mehrtash et al. (2017) the authors make use of mp-MRI (DW, DCE and T2w) at the patient level (3D), and extract 32x32 voxels by making use of the ROI. They add explicit zone information of the tumor to the first layers of the VGG-inspired architecture. They obtain an AUC of 0.80 for the test set. Finally, in (Yoo et al., 2019) DW MRI is used. They extract features after obtaining voxels from the aforementioned pictures and train a DL architecture. The features obtained from the DL model are then fused and used as an input for a decision tree. They obtain An AUC of 0.87 at the slice level and of 0.84 at the patient level. Similarly to the previous cases, the input of the network expects voxels extracted around the ROI, which which makes it hard to translate to other datasets or institutions without the ability to obtain such annotations.

Our work also relates to the work of (Shen et al., 2019), in which a TL approach exploiting mammogram patches is presented. In their work, an "end-to-end" approach to classify malignant tumors is presented. The authors experiment with different DL architectures and test the effect of adding more complexity in the form of new layers to the pre-trained model. The obtained results are encouraging, reaching an AUC of 0.95 at the slice level.

### 1.2. Contributions

In light of the difficulty to obtain annotations at the lesion level and the sub-optimal ImageNet initialization, we propose an approach that does not require ROI annotations at evaluation time and exploits patches extracted from the data to obtain a more optimal initialization method when compared with ImageNet one. The main contributions of this paper can be summarized as follows:

- We present a comprehensive comparison between different DL architectures and different settings for 2D slice-based PCa triage (types of loss, augmentation to the data, regularization and TL) that serves as a baseline for our work.

- We propose a classifier that exploits TL (self-TL) by making use of patches extracted from the MRI and does not require a lesion-annotated set nor a ROI extraction method in evaluation time.

- We explore the effect of TL based on ImageNet weights on our self-TL method and show the positive effects of our method when dealing with a limited amount of labeled data.

- We present a comprehensive ablation of our proposed method that includes the effect of number of patches, a preselection strategy for the patches and the effect of their location on the image as well as effect of different loss functions, augmentations and regularization of the network.

- We introduce a cross-TL method in which we pre-train the network making use of a single MRI modality as a source domain and transfer the weights to the other MRI modality (target domain).

## 2. Materials and Methods

A general scheme of the technical approach to the project is presented in Figure 2.

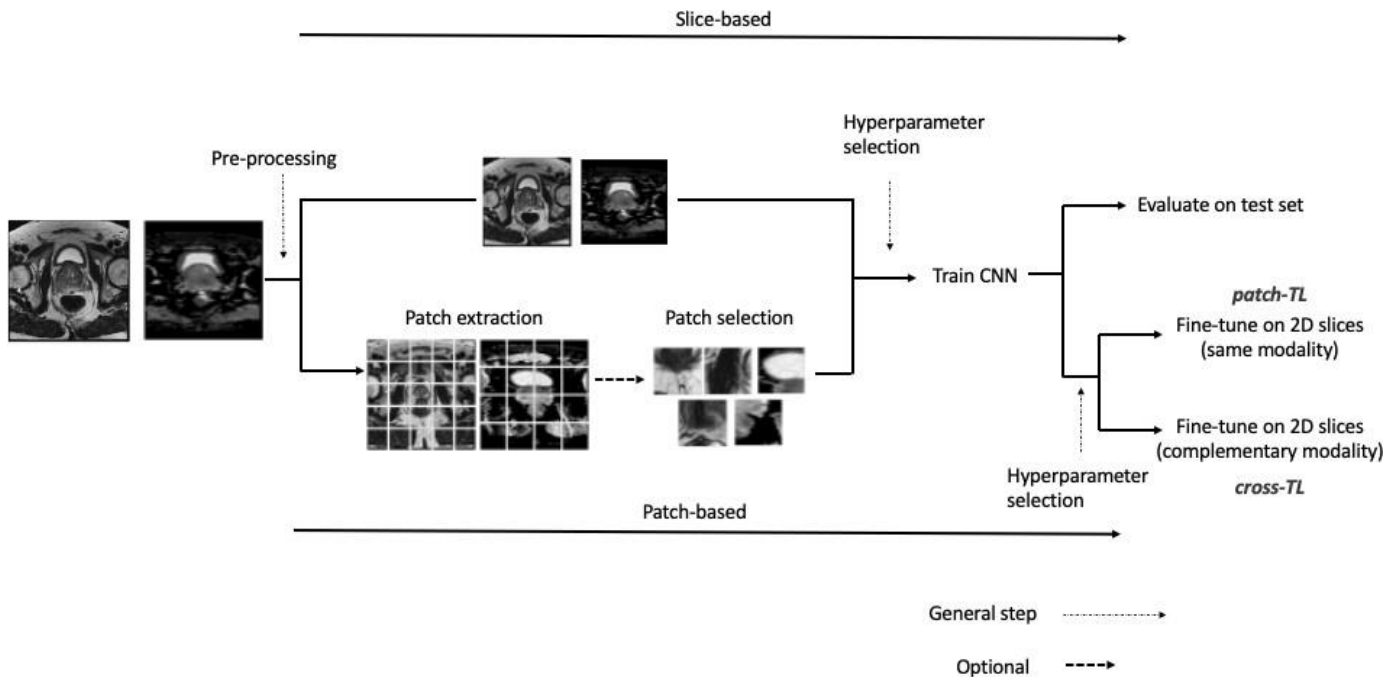

Fig. 2: Technical approach to the project.

## 2.1. Data

The data used for the development and validation of the models comes from the PROSTATEx challenge (Litjens et al., 2014), an open source data set[1]. The cohort included in the study consisted of 204 patients diagnosed with PCa and 330 lesions. Among those lesions, 76 lesions were cS and 254 were ncS. Lesions were annotated by experienced radiologists and assigned a PI-RADS score. Findings with a PI-RADS score 3 were referred to biopsy (Armato et al., 2018). The nature of the study is retrospective and includes different MRI modalities. We make use of ADC and T2w axial sequences, based on current PI-RADS guidelines.

The images were acquired on two different types of Siemens 3T MRI scanners. T2w were acquired using a turbo spin echo sequence and had an original resolution of 0.5 mm in-plane and a slice thickness of 3.6 mm. DW series were acquired with a single-shot echo planar imaging sequence with a resolution of 2 mm in-plane and 3.6 mm slice thickness. Three b-values were acquired (50, 400 and 800 s/mm²) and, subsequently, the ADC map was calculated by the scanner software. All images were obtained without an endorectal coil.

In addition to the data from the ProstateX challenge, we also make use of the the lesion masks (ROI) provided by Cuocolo et al. (2021)[2], obtained by two residents and two experienced board-certified radiologists. The lesion masks are used to label the bp-MRI in a slice-by-slice basis. Examples of both the original pictures and their respective masks are shown in Figure 1.

### 2.1.1. Pre-processing and data splitting

In order to homogenize the differences between the MRI of the dataset (namely, the different sequences and modalities) we perform different steps. The first step consists of re-sampling the MRI into a common coordinate system since the original resolution ranged from 320x320 to 640x640 and from 106x128 to 128x120 for T2w and ADC, respectively. We re-sample the sequences by linear interpolation to obtain a resolution of 320x320 and 128x128, for T2w and ADC, respectively. Following, we normalize the intensity of the sequences to an interval of [0, 1] and histogram cropping is performed by forcing the pixel intensity values of the sequences between the 1st and 99th percentiles.

We split the bp-MRI sequences in three different sets for each modality: training, validation and test. The splitting is performed in a stratified way, making sure the proportions with respect to the number of cS and ncS lesions in the original data set are kept. The splitting is performed by patient, avoiding at all costs possible cross-contamination in the form of lesion-spreading of the same patient across the different sets. The number of patients used for the study is 200 and the number of unique lesions among those patients is 299. From those 299 lesions, 76 are cS and the remaining 233 are ncS. We use 70% of the data to train, 20% to test the performance of the model and 10% to fine-tune the models and test the different model configurations, resulting in 864 slices to train (612 nCS and 252 cS), 221 to test (162 nCS and 59 cS) and 110 to validate and fine-tune (81 nCS and 29 cS). A summary of the patch-based approach statistics including number of patches per set and class distribution for train and test sets is presented in Table 2.

## 2.2. Methods

The work is divided in different methodological sections. In the first part of the work, we aimed to provide a comprehensive comparison between different CNN-based architectures and hyperparameter settings, including loss functions, regularization, use of TL (ImageNet) and use of augmentation for PCa triage: classification between ncS and cS lesions. In particular, the first part of the methodology describes the process to obtain results based on slices (full image, 2D - Figure 2, upper path), following a standard procedure: train on the MRI slices contained in the training set, find the best hyperparameter setting using the validation set and report results on the test set with the best configuration found with the validation set. We provide details about the different configurations tested to reach what serve as baselines for following comparisons with our proposed approaches.

In the second part, we describe the first of our proposed approaches. In particular, we describe the process followed to obtain patches from the original MRI slices which were used to train and validate the baselines, and how we translate the training process from the source domain (patch-based) to the target domain (slice-based, Figure 2 - lower path). In addition, we provide details on the different settings tested for the pre-training strategy: pre-selection of patches, loss functions, augmentation and TL.

In the third point, we present the methodology followed for our cross-TL approach in which we leverage the pre-training done in a specific modality domain (namely, T2w) and we transfer it to another modality target domain (namely, ADC). In this particular setting, we transfer, again, features obtained from a patch-based domain to a full-slice domain (Figure 2, lower path). We provide details on the different testing done such as the number of frozen layers in the architecture when performing the cross-TL approach. Finally, the metrics used to evaluate all the methodology are presented.

### 2.2.1. Slice-based triage

In the first part of the work we focus on classification between cS and ncS at the slice level. The input for our DL architectures is the pre-processed MRI slice (Section 2.1.1) of either

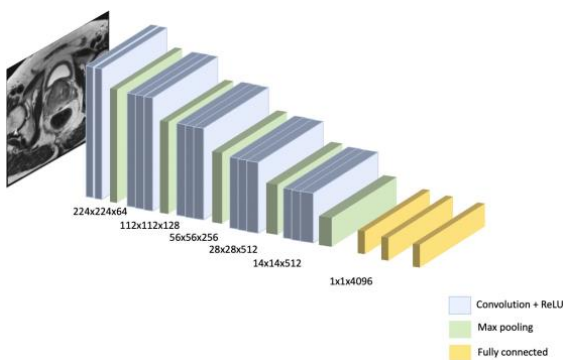

Fig. 3: Whole slice methodology based on T2w. For the sake of simplicity we assume a VGG16 architecture.

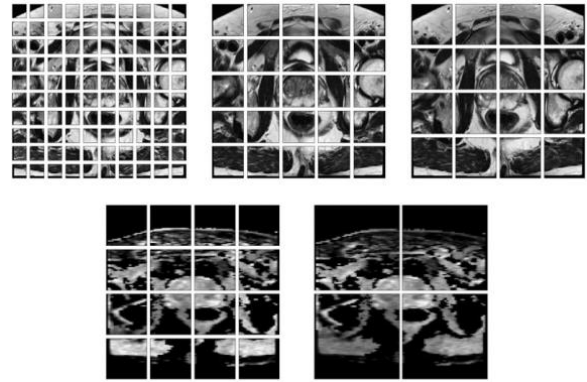

Fig. 4: Result of dividing the different MRI modalities in patches of different sizes. Top row: T2w (32by32, 64by64 and 80bby80). Bottom row: ADC (32by32 and 64by64).

T2w or ADC modality. We label the data on a slice-by-slice basis where the label $y = \{0, 1\}$ depending on whether the slice under consideration has a cS lesion ($y = 1$) or not (Figure 3). Methods are applied in the same way for each MRI modality without taking into account the combination of both, considered to be out of the scope of the article.

***Architectures:*** Different CNN architectures are tested and compared on the basis of the final AUC obtained on the test set. Specifically, we focus on the following architectures: VGG16 , VGG19 (Simonyan and Zisserman, 2014), ResNet18, ResNet34 (He et al., 2016) and GoogLeNet V3 (Szegedy et al., 2016), based on the most commonly used architectures in the different works in PCa triage (Section 1.1). We experiment with both regularized architectures and non-regularized. We use $L_2$ regularization and experiment with different values: $1e^{-4}$, $1e^{-5}$ and $1e^{-6}$.

***Loss:*** We experiment with two loss functions. The first one is the commonly used *binary cross entropy* for classification tasks (Eq. 1).

$$L(y_i) = \frac{1}{N} \sum_{y_i=1} y_i log(p(y_i)) + (1 - y_i)log(1 - p(y_i)) \quad (1)$$

In addition to it, we also make use of *focal loss* (Lin et al., 2017). The rationale underlying the chose of focal loss is its ability to cope with imbalanced data sets. In essence, focal loss is an extension of cross-entropy which takes into account imbalance present in the data by down-weighting the most represented class samples (with a factor $\gamma$) as well as by adding a multiplicative factor $\alpha$ which is meant to bring attention to the less represented class (Eq. 2). We fix both parameters during the training of the networks to $\gamma = 2.0$ and $\alpha = 0.25$, based on literature results.

$$L(p, y) = \begin{cases} -\alpha(1 - p)^\gamma log(p) & y = 1 \\ -(1 - \alpha)p^\gamma log(1 - p) & otherwise \end{cases} \quad (2)$$

***Training parameters:*** We fix the batch size to 64 for T2w sequences and to 128 for ADC, due to graphical memory capacity and that makes use of the full image in the patch extraction process. In addition, we explore the effect of an offline patch-selection process in which we filter the total number of patches extracted from the image and keep only the ones deemed as relevant based on prior knowledge about prostate zones and the prevalence of tumors. In order to train the network, we label the patches based on the label of the original slice (i.e. y={0, 1}, where 0 is ncS and 1 is cS). During training, we first train the network on the obtained patches (source domain). During that process, we explore several configurations such as TL (including layer freezing level), augmentation, loss function and regularization. Once the network has been trained and the hyperparameters have been chosen, we re-train the network on the slice domain (target domain) by making use of TL. Such process allows us to to have a slice-based triage system which does not require ROI annotations during evaluation time. We apply the aforementioned methodology to both MRI modalities in an independent way.

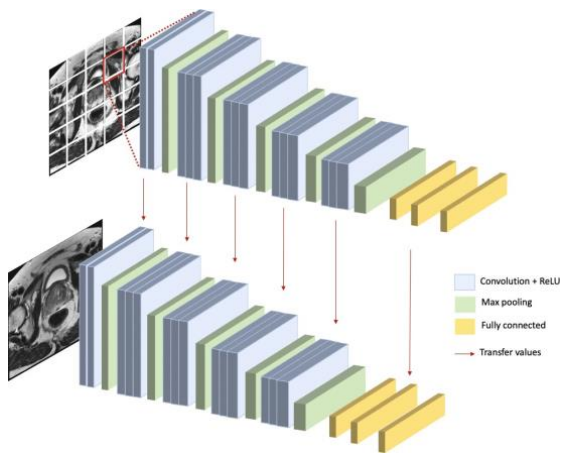

Fig. 5: Exemplification of our patch-based TL methodology based on T2w. For the sake of simplicity we assume a VGG16 architecture and that the pre-trained network is trained on only the source domain and no ImageNet weights are involved.

*Architectures:* We hypothesize the best architecture in terms of validation AUC found in the slice-based triage will also be the most suitable one for the patch-based setting due to the similar nature of the data. Hereby, in this case, we fix the architecture and make use of VGG16 based on the previous experiments. We explore different $L_2$ regularization values: $1e^{-4}$, $1e^{-5}$ and $1e^{-6}$.

*Loss:* Similarly to the slice-based experiments, we investigate the effect of *binary cross entropy* (Eq.1) and *focal loss* (Eq.2) on the source domain.

*Training parameters:* We fix the batch size to 1024 for ADC sequences and to 2048 for T2w ones, due to graphical memory capacity limitations. We experiment with the same learning rates as for the slice-based case: $1e^{-4}$, $1e^{-5}$ and $1e^{-6}$. We train for a fixed number of epochs (5000) and make use of an early stopping mechanism on the validation set in which if the validation loss does not have a significant improvement over 40 iterations the training is halted.

*Augmentation:* We investigate the effect of augmentation in the pre-training phase and in the target domain one. We select augmentation techniques based on previously published works Le et al. (2017); Hao et al. (2020), consisting of rotation (-50 to 50 degrees), translation (range of 0.32) and vertical flipping.

*Transfer Learning:* We investigate the effect of TL based on ImageNet weights on the pre-training task in the patch domain.

limitations. We experiment with different learning rates: $1e^{-4}$, $1e^{-5}$ and $1e^{-6}$ for both sequences and for each possible configuration. We train for a fixed number of epochs (2000) and we use an early stopping mechanism on the validation set in which if the validation loss does not have a significant improvement over 40 iterations the training is halted.

*Augmentation:* We investigate the effect of augmentation in the PCa triage application. We select augmentation techniques based on previously published works (Le et al., 2017; Hao et al., 2020) consisting of rotation (-50 to 50 degrees), translation (range of 0.32) and vertical flipping.

*Transfer learning:* We test the effect of TL on the final PCa triage AUC. We make use of ImageNet weights and experiment freezing different layers of the networks. Since the original data set is gray-scale, it needs to be transformed to be suitable for the TL setting due to the RGB nature of ImageNet weights. We apply a simple transformation to the original data in which the slice is copied over the other channels, making it suitable for the TL setting. We start the TL process by training the network with a small learning rate (i.e. $1e^{-8}$) to let the network adapt and continue the process with a higher learning rate (i.e. previously mentioned values) after a few epochs.

### 2.3. Patch-based self-Transfer Learning (self-TL)

In our first proposed method, we obtain patches from the MRI sequences to increase the amount of data as well as to allow the network to capture more fine details at the lesion resolution. In order to obtain the patches, we process every slice with a sliding window with predefined size and without overlap nor without offset in between the different patches.

The patches are obtained in an offline way (i.e. before the training) and we explore different patch sizes: 32x32 and 64x64 for ADC and 32x32, 64x64 and 80x80 for T2w (Figure 4). The minimum size is selected based on the DL architecture requirements. As for the other dimensions, we look for sizes that allow a patch extraction process that does not require padding

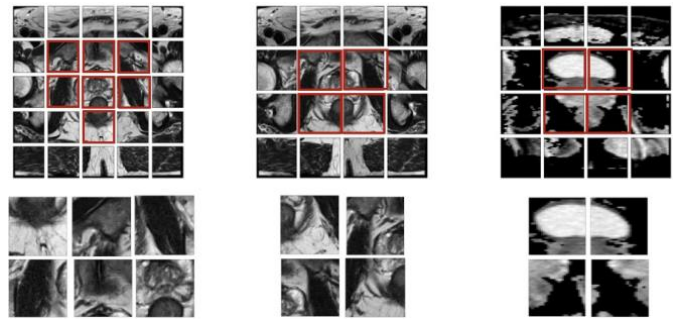

Fig. 6: Example of pre-selection of patches based on tumor frequency location.

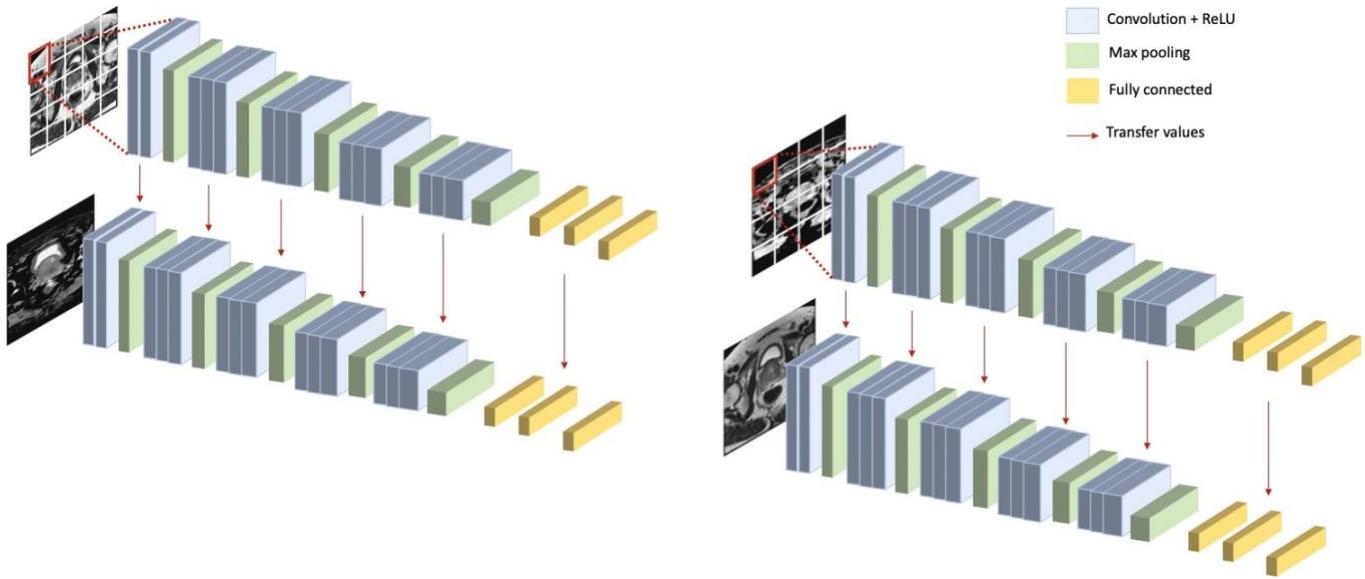

Fig. 7: Cross-modality TL. Left side: Transfer learning from T2w domain to ADC target domain. Right side: Transfer learning from ADC domain to T2w target domain. For the sake of simplicity a VGG16 architecture is assumed.

Similarly to the slice-based case, the patches are gray-scale, thus not suitable for TL based on ImageNet in their original form. We modify the data in the same way as the slice case: copying the patch over the other channels. We start the process by training the network with a small learning rate for a few epochs and then continue the training process with a higher learning rate. If TL is used in the pre-training phase, the slice (target domain) also needs to be modified. We apply the same methodology: copy the slices over the other channels.

***Self-Transfer Learning:*** Once the DL architecture has been trained on the patch domain we manipulate it to be able to use it at the slice level. In particular, we reuse the weights obtained from the patch training process while changing the input size of the network to a suitable one for the sequence under consideration (320x320 T2w and 128x128 ADC). We experiment with different training parameters in the target domain. Specifically, we investigate the effect of the previous proposed loss functions (Eq. 1, 2) and the usage or non-usage of previously proposed augmentation techniques. We apply $L_2$ regularization and fix it to a value of $1e^{-5}$. We use a batch size of 64 for T2w and 128 for ADC.

***Patch selection:*** Different approaches are explored in the pre-training based on patches (source). The first approach is to train with all the patches extracted from every slice. The second approach considers a pre-selection of patches on the basis of the most common locations of the PCa lesions. Specifically, we pre-select patches around the gland of the prostate for every slice (Figure 6). The number of pre-selected patches depends on the size of the extracted patches. Table 2 presents a summary of the number of extracted patches and the different sizes tested in our approach.

### 2.4. Cross-modality Transfer Learning

In our second proposed method, instead of considering both MRI modalities in an independent way, we consider the ef-fect of making use of both of them to improve the previous results. As such, we tackle the triage problem with a patch-based pre-training approach, based on the results obtained in the previous section. However, in this case, instead of pre-training with patches extracted from a single modality and transferring the weights to the slice domain of the same modality we pre-train with patches extracted from one modality and transfer the weights to the slice domain of the complementary MRI modality (Figure 7). The intuition behind our proposed methodology is that current PI-RADS guidelines suggest that the main sequence for PCa triage is ADC, since a high % of the tumors are located in the peripheral zone of the prostate. Hereby, intuitively, we try to extract features from the "support" modality while making the final triage decision on the modality that is supposed to be able to detect most of the tumors, imitating (with obvious differences) the radiologists choice when it comes to PI-RADS-based triage.

***Training:*** The choice of the architecture is based on empirical results obtained after experimenting with the different architectures: VGG16. The pre-training based on patches uses the same configuration as the one presented in the previous section for ADC and T2w. As for the target domain (once the weights from the pre-trained network have been transferred (i.e. ADC or T2w slices): we investigate the effect of the loss function. We fix the $L_2$ regularization value to $1e^{-4}$. We use a learning rate of $1e^{-5}$ and an early stopping mechanism of 40 iterations if no improvement is seen in the validation loss. We experiment with different pre-training settings based on the different proposed patch sizes and patch selection strategy, as explained in the previous self-TL approach (Section 2.3). In addition, we also explore different layer freezing levels. After pre-training in the patch domain, we transfer the weights to the slice-based domain and progressively train the network with small learning rates (order of 1e-7) in order to make sure the network adjusts to the new domain while still making use of the learned features.

Table 2: Sizes of sets when using patch-based approach and selection of patches.

| Modality | Set | Patch size | Selection | Number of patches | ncS patches | cS patches |
|---|---|---|---|---|---|---|
| T2w | Train | 80x80 | C | 13824 | 9792 | 4032 |
| | | 64x64 | C | 21600 | 15300 | 6300 |
| | | 32x32 | C | 86400 | 61200 | 25200 |
| T2w | Train | 80x80 | √ | 3456 | 2448 | 1008 |
| | | 64x64 | √ | 5184 | 3672 | 1512 |
| | | 32x32 | √ | 15552 | 11016 | 4536 |
| T2w | Test | 80x80 | C | 3536 | 2592 | 944 |
| | | 64x64 | C | 5525 | 4050 | 1475 |
| | | 32x32 | C | 22100 | 16200 | 5900 |
| T2w | Test | 80x80 | √ | 884 | 648 | 236 |
| | | 64x64 | √ | 1326 | 972 | 354 |
| | | 32x32 | √ | 3978 | 2916 | 1062 |
| ADC | Train | 64x64 | C | 3456 | 2448 | 1008 |
| | | 32x32 | C | 21600 | 15300 | 6300 |
| ADC | Train | 32x32 | √ | 3456 | 2448 | 1008 |
| ADC | Test | 64x64 | C | 884 | 648 | 236 |
| | | 32x32 | C | 5525 | 4050 | 1475 |
| ADC | Test | 32x32 | √ | 884 | 648 | 236 |

All of our experiments are run using Keras (Chollet et al., 2015). If not stated otherwise, the following applies to all the experiments and methodology: we keep the best weights for every experiment based on the validation set AUC values per epoch. We use Adam (Kingma and Ba, 2014) optimizer. All of the experiments are run on a Tesla V100 with 30GB of RAM.

### 2.5. Evaluation of results: metrics

In this section we introduce the different metrics employed to evaluate the results:

***Sensitivity and specificity:*** It measures the ratio of cS lesions (True positives - TP) that are classified as cS lesions with respect to the total number cS lesions (miss-classified and correctly classified). For instance, a sensitivity of 65% would imply that 65% of the cS lesions were classified as cS but a 35% of them were miss-classified. Similarly to the sensitivity, the specificity measures the number of ncS lesions (True negatives - TN) that are classified as ncS with respect to the total number of ncS lesions (miss-classified and correctly classified). Classification models typically associate a confidence score to each prediction (i.e.: a value from 0 to 1), which can be used to tune the sensitivity-specificity interrelationship at will, by adjusting the confidence threshold, denoted as $t$ in Eq. 3. For example, by setting the threshold to 0.01, almost all lesions will be considered clinically significant, leading to very high sensitivity yet very low specificity; while, by setting the threshold to 0.99, the opposite would happen.

$$Sensitivity@t = \frac{TP}{TP + FN} \quad Specificity@t = \frac{TN}{TN + FP} \quad (3)$$

***Area Under the Receiving Operating Curve (AUC):*** A summary of sensitivity and specificity for all possible threshold $t$ values. Is computed as the area under the curve that emerges by plotting the sensitivity against 1-specificity.

## 3. Results

In this section, we provide the results obtained using the different methods proposed in the article, as well as the ones considered to be the baseline for our work. We start by showing the different results deemed as baseline in our work and used to have a fair and standardized comparison between the different methodologies based on ProstateX public data set. Baseline results are slice-based for both modalities (ADC and T2w). Following, we present the results based on our self-TL proposed patch approach and compare them against the baseline ones. Finally, cross-TL results are presented based on the previous patch approach and compared against the other two approaches. In addition, we show how our proposed methods outperform ImageNet and random initialization as well as the robustness of our proposed methods when dealing with data scarcity. Metrics used to evaluate the different results can be found in section 2.5.

### 3.1. Slice-based triage

A comprehensive evaluation of the results obtained for T2w and ADC modalities is presented in Table 3. For the sake of simplicity, we present the best results based on AUC for each architecture. In particular, we report the test AUC as the standard metric to compare our results.

***Experimental setup*** As section 2.2.1 introduced, we experiment with different hyperparameters configurations. The tables show the best configurations out of all the tested ones. The best configurations are chosen based on the validation set whilst the final results are based on test set. We experiment with different architectures, loss, number of frozen layers, regularization and augmentation. We train the chosen architecture for a defined number of epochs and apply early stopping if the loss does not improve in 40 consecutive epochs. Once the best hyperparameters have been chosen, we test the best configuration on an independent test set.

Table 3: Slice-based results for T2w modality with the best set of hyperparameters out of all the tested configurations.

| Modality | Architecture | Epochs | BS[*] | LR[*] | Regularization | Loss | TL[*] | Frozen layers | Augmentation | AUC[t] |
|---|---|---|---|---|---|---|---|---|---|---|
| | VGG16 | 2000 | 64 | $1e^{-6}$ | $L_2$: $1e^{-4}$ | CE[*] | √ | 15 | C | **0.741** |
| | VGG19 | 2000 | 64 | $1e^{-6}$ | $L_2$: $1e^{-4}$ | CE | √ | 20 | C | 0.733 |
| T2w | ResNet18 | 2000 | 64 | $1e^{-6}$ | $L_2$: $1e^{-4}$ | CE | C | C | C | 0.659 |
| | ResNet34 | 2000 | 64 | $1e^{-6}$ | $L_2$: $1e^{-4}$ | CE | √ | 158 | C | 0.680 |
| | Inceptionv3 | 2000 | 64 | $1e^{-6}$ | $L_2$: $1e^{-6}$ | FL[*] | C | C | C | 0.714 |
| | VGG16 | 2000 | 128 | $1e^{-5}$ | $L_2$: $1e^{-5}$ | FL[*] | √ | 15 | √ | **0.785** |
| | VGG19 | 2000 | 128 | $1e^{-5}$ | $L_2$: $1e^{-6}$ | FL | √ | 20 | √ | 0.704 |
| ADC | ResNet18 | 2000 | 128 | $1e^{-5}$ | $L_2$: $1e^{-4}$ | CE[*] | C | C | √ | 0.692 |
| | ResNet34 | 2000 | 128 | $1e^{-5}$ | $L_2$: $1e^{-6}$ | CE | √ | 147 | C | 0.646 |
| | Inceptionv3 | 2000 | 128 | $1e^{-6}$ | $L_2$: $1e^{-4}$ | CE | √ | 299 | √ | 0.637 |

[*] BS = Batch size, LR = Learning rate, TL = Transfer learning, CE = Cross-entropy and FL = Focal loss.

[t] Test AUC using the set of hyperparameters obtained through the validation data.

**Results** The best architecture for both modalities is **VGG16**, which is coherent with most of the literature results. In the case of ADC, the best configuration is **FL loss with TL, 15 frozen layers and augmentation**. On the other hand, T2w obtains the best results with **CE loss with TL, 15 frozen layers and no augmentation**. The best results were an **AUC of 0.741 for T2w** and an **AUC of 0.785 for ADC**, both obtained with the previously mentioned VGG16.

**Discussion** Influence of loss functions on the final results are slightly different for both modalities: FL appears to have a positive influence on final AUC results for ADC modality. On the other hand, CE improves around 1% the results for T2w modality when compared to FL. We hypothesize that since ADC is the predominant sequence for tumor located in the peripheral zone and there is a larger shift in the lesion distribution towards ncS in such location, the imbalance on the location might make FL loss work better with ADC. Regularization had a positive effect on both modalities, avoiding an early overfitting to the data under consideration after a few epochs of training. Its positive effects can, presumably, be attributed to the limited size of the data set when using the full slice. We experimented with a different number of frozen layers when considering a pre-trained architecture on ImageNet. In both cases, we observed a positive effect on the final results when considering such setting and only unfreezing the last layers of the model, which are supposed to be representative of the less generalizable features of the data under consideration. In terms of augmentation, we observe how T2w data does not benefit from it. On the other hand, ADC clearly benefits from augmentation with 4 out of the 5 architectures obtaining the best results with its application. The differences between modalities might indicate that our set of augmentation techniques might be better suited for one modality but not for the other one. Finally, we can see how ADC has a better discrimination ability than T2w. We can associate such ability to the fact that ADC is the predominant sequence in PCa triage.

### 3.2. Patch-based triage

Table 4 presents the results obtained for both modalities following the patch-based approach introduced in section 2.3.

Specifically, we show the best results in terms of test AUC for every patch size we experimented with once the best hyperparameters were found with the validation set.

**Experimental setup** All the patches were extracted without overlap and without padding in order to avoid redundancy among the extracted patches. The minimum size of the patches is limited to 32x32 due to inherent limitations of VGG16 but also, because we hypothesize that smaller patches might not carry enough discriminating diagnostic information. We experiment with patches that are considerably large compared to the typical size of a tumor given that they have a larger fiew of vield and provide a larger discriminating ability 4.The number of patches obtained for each modality is depicted in Figure 4. Finally, we experiment with a patch-selection approach based on frequency location of the tumors (Figure 6. We hypothesize that patches around the ROI (that is, located close to the tumor - peripheral, transition or anterior fibromuscular zones) might carry more discriminative information. Hereby, we perform an automatic selection of patches based on patch proximity to the ROI. We test the same hyperparameter configuration as in the slice-based approach, selecting the best configuration based on the validation set.

**Results** We observe how the **best AUC is obtained with 32x32 patches and no pre-selection for T2w**. However, results obtained with patches of size 80x80 and pre-selection are quite close to the best ones. In particular, we obtain an **AUC of 0.886** and 0.879, respectively. Training for the patch-based pre-training is performed from scratch without transferring ImageNet weights. As for ADC, best results are obtained again, with 32x32 patches and a pre-selection strategy which selects 4 patches per slice. Training for the patch-based pre-training is again performed from scratch, without making use of ImageNet weights. In particular, best **AUC results are 0.831**. We can see how in both cases there is a significant improvement when compared to the slice-based approach (0.741 vs 0.886 for T2w and 0.785 vs 0.831 for ADC), being T2w sequence the most benefited one from the patch approach. In both cases the final classifier struggles with the less represented class (cS cases) as depicted by the sensitivity@0.5.

Table 4: Patch-based results for T2w and ADC modalities with the best setting combination out of all the ones tested and VGG16 architecture.

| Modality | Size | Selection | Number selection | TL* | Frozen ImageNet | Pre-augm.* | Frozen self-TL* | Loss | Augm.* | Sensitivity@0.5 | Specificity@0.5 | AUC† |
|---|---|---|---|---|---|---|---|---|---|---|---|---|
| T2w | 80x80 | ✓ | 4 | C | C | ✓ | 15 | FL* | ✓ | 0.831 | 0.927 | 0.879 |
|  | 64x64 | ✓ | 6 | C | C | ✓ | 15 | CE* | ✓ | 0.706 | 0.762 | 0.794 |
|  | 32x32 | C | C | C | C | C | 15 | CE | ✓ | 0.819 | 0.931 | **0.886** |
| ADC | 64x64 | C | C | ✓ | 15 | C | 15 | CE | ✓ | 0.820 | 0.987 | 0.817 |
|  | 32x32 | ✓ | 4 | C | C | C | 15 | CE | C | 0.841 | 0.910 | **0.831** |

* TL = Transfer learning, Pre-augm. = Pre-training augmentation, Frozen self-TL = Frozen layers self-Transfer Learning,
Augm. = Augmentation, FL = Focal loss, CE = Cross-entropy.

† Test AUC.

Table 5: Cross-modality approach results for T2w and ADC modalities with the best setting combination out of all the ones tested and VGG16 architecture.

| Source | Target | Size | Selection | Number selection | TL* | Frozen ImageNet | Pre-Augm.* | Frozen layers cross-TL* | Loss | Augm.* | Sensitivity@0.5 | Specificity@0.5 | AUC† |
|---|---|---|---|---|---|---|---|---|---|---|---|---|---|
| T2w | ADC | 80x80 | ✓ | 4 | C | C | C | 11 | FL* | ✓ | 0.820 | 0.912 | 0.862 |
|  |  | 64x64 | ✓ | 6 | C | C | ✓ | 11 | FL | ✓ | 0.845 | 0.940 | **0.898** |
|  |  | 32x32 | C | C | C | C | ✓ | 11 | CE* | ✓ | 0.833 | 0.916 | 0.876 |
| ADC | T2w | 64x64 | C | C | C | C | ✓ | ✓ | FL | ✓ | 0.661 | 0.904 | 0.815 |
|  |  | 32x32 | C | C | C | C | C | 11 | CE | ✓ | 0.624 | 1.000 | **0.837** |

* TL = Transfer learning, Pre-augm. = Pre-training augmentation, Frozen cross-TL = Frozen layers cross-Transfer Learning,
Augm. = Augmentation, FL = Focal loss, CE = Cross-entropy.

† Test AUC.

***Discussion*** Different number of patches are tested and the best results (Selection and number columns in Table 4) are shown in Figure 6. We empirically verify that VGG16 is the best suited architecture for the task. Following, we test different configurations and validate them through the validation set. In particular, we find out that the re-training on the full slices benefits from a pre-training strategy based on the patches without TL in the vast majority of cases. We hypothesize that might be due to the gap between the domains, being the patch-based one closer to the target domain (slices) and hereby, being more optimal as a warm-up strategy. In addition, we also experiment applying augmentation in the patch pre-training strategy. Results show that in some cases, the pre-training strategy benefits from it (see T2w modality in Table 4). The reasoning as why that is the case it is not as clear. Further experiments would need to be carried out in order to comprehend the effect of the different augmentation strategies and which ones are more optimal for each domain, which is out of the scope of this paper. Furthermore, we experiment with ImageNet initialization in the patch domain. Such initialization did not have a positive effect, showing that our in-domain pre-training strategy is more optimal for the task under consideration when compared to it.

Finally, we experiment with the automatic patch selection around the ROI. Our results show that the approach clearly benefits from it. We justify its positive influence on the amount of discriminative information carried out by them and that manually ruling out those patches that might not contain any useful information might help the learning process of the network. In particular, we can see that the patch selection strategy is not as effective with 32x32 patches in the T2w case and with 64x64 patches in the ADC one. We hypothesize it might be due to the fact that those resolutions carry the least discriminative information for each MRI modality, since 32x32 is relatively small and patches barely contain any relevant prostate parts and the 80x80 ones divide the ADC sequence in non-coherent sections.

### 3.3. Cross-modality triage

Finally, we present the results for our cross-modality TL approach in table 5. In a similar fashion to the previous section, we make use of patches to pre-train a VGG16 architecture and then fine tune it on the whole slice. In this case, we pre-train using patches from one source modality and fine-tune on the opposite target modality. Intuitively speaking, we follow the radiologist process in which one of the modalities serves as a support for the other one when evaluating patients' MRI.

***Experimental setup*** In this case, we train the architecture on a defined patch domain (either ADC o T2w) and then use TL and re-train the architecture on the slice domain of the opposite modality. Again, we evaluate the proposed method in different scenarios (Table 5): different patch size (Size), the proposed patch selection approach (Selection), number of pre-selected patches (number selection) and fine-tuning ImageNet weights during the patch training in the source domain as well as different frozen layers if used. In addition to it, we also experiment freezing a variety of number of layers in the target domain (Frozen layers cross-TL), after transferring the weights obtained through the pre-training in the target domain (patch-based). We present the best results for each patch size based on the evaluation set selected hyperparameters. The re-training in the target domain (slice) consists of a progressive training in which we begin with low learning rates (order of $1e^{-8}$) to allow the network to adapt the weights to the new modality. Following, we carry out a normal training using a higher learning rate once the network has adapted to the target domain (order of $1e^{-5}$). In essence, we could consider such process as a domain adaption one.

***Results*** We obtain the best results for ADC as a target domain and T2w as a source one, with a 64x64 patch-based configuration with pre-selection for the pre-training strategy. The **AUC is of 0.898** on the test set with a **sensitivity of 0.845** and **specificity of 0.940**, for a threshold of 0.5. As for using ADC as the source domain and fine-tuning on T2w, we obtain an **AUC of 0.837**, **sensitivity of 0.624** and **specificity of 1.000** for a threshold of 0.5 and patches of size 64x64 without pre-selection.

*Discussion* As shown in table 5 the best results are obtained by pre-training on the T2w domain and using the ADC one as the target domain. In particular, in the T2w case we observe how the results surpass the previously presented results based on a single modality and patch-based pre-training for both modalities. In the ADC case, results are slightly better than the ones obtained training solely on ADC but worse than the ones obtained in the target domain (T2w). Similarly to the previous results, we observe how the classifier struggles with the less represented class (cS), as shown by the sensitivity values. Interestingly, we obtain the best results after applying a patch selection approach which again, we hypothesize might be due to the amount of useful information for the classification task the patches carry. Finally, once again we get the best results when the pre-training based on patches is done without ImageNet weights, which shows once again that our in-domain method is more optimal as an initialization strategy than ImageNet one. In addition, we observe how the best results are obtained when T2w is used as a source rather than a target. Intuitively speaking, we presume that would be because T2w carries more structural details of the prostate thus benefiting more from the patch selection approach as well as providing the network more information in the lowest layer levels of the network.

## 4. Conclusions and remarks

In this paper, we proposed a novel initialization method based on patches and a cross-TL one to make use of bi-parametric data in a similar way radiologists' approach PCa triage. In addition, we presented a comprehensive comparison between an extensive list of different settings and DL architectures which we further used as baselines for following comparisons.

Our patch-based approach enables a VGG16 architecture to extract better diagnostic-discriminating features by extracting blocks with sizes close the typical tumor size. Thanks to the TL nature of our approach we are also able to avoid ROI labels on subsequent target domain whilst still obtaining a high AUC for PCa triage. We have validated our proposed patch-based methodology using 2D bi-parametric prostate MRI and observed how the proposed methodology outperforms the proposed baselines. In addition, we provide a comprehensive comparison between different patch sizes and settings in which we pre-select patches based on prior zonal knowledge of tumor locations.

Our cross-TL enables a VGG16 architecture to benefit from both modalities without any additional complexity in the form of added modules to the original CNN architecture. We have also validated our proposed methodology using both T2w and ADC as the source domain whilst using the remaining sequence as the target one. We provide a comparison between different patch sizes and selection strategies as well as other settings such as the number of frozen layers when fine-tuning in the target domain. We observe how in some cases a selection strategy based on frequency of appearance of the tumor in that specific zone is useful to learn more discriminative features, as the selection allows to focus on patches that are likely to carry them.

In both proposed methodologies we observe how we consistently outperform the proposed baselines by a large margin. Moreover, we reach AUC levels that are on pair with the ones presented in the related work but with several advantages over the other proposed methodologies such as making use of less MRI modalities (bi-parametric based in our case) which might not be available and are expensive to label as well as not requiring ROI annotations during evaluation time. In addition, our cross-TL method makes use of both modalities without any additional architecture complexity by leveraging TL. Moreover, we believe our cross-TL approach is intuitive from a clinical point of view which might result in a warmer welcome when presented in a real world scenario.

Our work has several limitations. First of all, we evaluate all the methodology on a single data set (open source) but further evaluation would be required on an external data set of another institution to check for possible data biases. Furthermore, the nature of the data is retrospective as well as of the study. Prospective studies are not common in the medical computer vision community but still of much need to validate results and bringing conceptual applications into practice. Hereby, we believe a prospective study would be interesting to enhance the quality of the presented results. Finally, and in spite of our cross-TL being successful, we believe that a pre-domain adaptation technique could be explored to enhance results even more.

Nowadays, many CNNs are pre-trained before the main task but the differences between the domains of the source and the target data might not allow to make use of the full potential of the pre-training strategy. To that end, self-supervised learning (SSL) aims to learn from unlabeled data to then fine-tune those learned representations in the target domain, which is commonly the same as the source one, thus making the gap between both domains smaller or non-existent. Our approach is closely related to SSL approaches in the sense that we try to close the gap between the domains. The promising results open the door to further exploration of SSL-based approaches for bi-parametric prostate MRI in which labels are completely omitted, following state of the art and trending tendencies in the computer vision field.

**Declaration of Competing Interest**
None.

**Acknowledgements**
This work has been funded by the University of Stavanger. The authors have no relevant financial or non-financial interests to disclose.